\documentclass[sigconf,natbib=true,anonymous=false,review=false]{acmart}
\usepackage{multicol}
\usepackage{multirow}
\usepackage{amsfonts}
\usepackage{bm}
\usepackage{amsmath}
\usepackage{amsthm}
\usepackage{colortbl}
\usepackage{fixltx2e}
\usepackage{xcolor}

\AtBeginDocument{%
  \providecommand\BibTeX{{%
    \normalfont B\kern-0.5em{\scshape i\kern-0.25em b}\kern-0.8em\TeX}}}

\copyrightyear{2024} 
\acmYear{2024} 
\setcopyright{acmlicensed}\acmConference[CIKM '24]{Proceedings of the 33rd ACM International Conference on Information and Knowledge Management}{October 21--25, 2024}{Boise, ID, USA}
\acmBooktitle{Proceedings of the 33rd ACM International Conference on Information and Knowledge Management (CIKM '24), October 21--25, 2024, Boise, ID, USA}
\acmDOI{10.1145/3627673.3679939}
\acmISBN{979-8-4007-0436-9/24/10}

\begin{document}

\title{How to Leverage Personal Textual Knowledge for Personalized Conversational Information Retrieval}

\author{Fengran Mo}
\orcid{0000-0002-0838-6994}
\affiliation{%
  \institution{RALI, Université de Montréal}
  \city{Montréal}
  \state{Québec}
  \country{Canada}
}
\email{fengran.mo@umontreal.ca}

\author{Longxiang Zhao}
\orcid{0009-0008-6503-3459}
\affiliation{%
  \institution{Dalian University of Technology}
  \city{Dalian}
  \state{Liaoning}
  \country{China}
}
\email{longxiang.zhao.1@gmail.com}

\author{Kaiyu Huang}
\orcid{0000-0001-6779-1810}
\affiliation{%
  \institution{Beijing Jiaotong University}
  \city{Beijing}
  \country{China}
}
\email{kyhuang@bjtu.edu.cn}

\author{Yue Dong}
\orcid{0000-0003-2161-8566}
\affiliation{%
  \institution{University of California, Riverside}
  \city{Riverside}
  \state{California}
  \country{United States}
}
\email{yue.dong@ucr.edu}

\author{Degen Huang}
\orcid{0000-0002-8860-7805}
\affiliation{%
  \institution{Dalian University of Technology}
  \city{Dalian}
  \state{Liaoning}
  \country{China}
}
\email{huangdg@dlut.edu.cn}

\author{Jian-Yun Nie}
\orcid{0000-0003-1556-3335}
\affiliation{%
  \institution{RALI, Université de Montréal}
  \city{Montréal}
  \state{Québec}
  \country{Canada}
}
\email{nie@iro.umontreal.ca}

\renewcommand{\shortauthors}{Fengran Mo et al.}

\begin{abstract}
Personalized conversational information retrieval (CIR) combines conversational and personalizable elements to satisfy various users' complex information needs through multi-turn interaction based on their backgrounds. The key promise is that the personal textual knowledge base (PTKB) can improve the CIR effectiveness because the retrieval results can be more related to the user's background. However, PTKB is noisy: not every piece of knowledge in PTKB is relevant to the specific query at hand. In this paper, we explore and test several ways to select knowledge from PTKB and use it for query reformulation by using a large language model (LLM). The experimental results show the PTKB might not always improve the search results when used alone, but LLM can help generate a more appropriate personalized query when high-quality guidance is provided.
  
\end{abstract}





\maketitle

\section{Introduction}
Personalized conversational information retrieval (CIR) systems aim to integrate personalized elements with conversational history, satisfying users' complex information needs through multi-turn interaction. Compared with previous CIR systems~\cite{qu2020open,yu2021few,mao2022convtrans,mao2022curriculum,mo2023learning,mao2023learning,jin2023instructor,mo2024history,mo2024convsdg,mao2024chatretriever}, it not only needs to address context-dependent queries to reveal the real users' information needs but also has to consider their backgrounds when searching the candidate documents.

A common practice to achieve CIR is leveraging conversational query reformulation techniques to transform context-dependent queries into stand-alone ones~\cite{yu2020few,voskarides2020query,lin2021contextualized,vakulenko2021question,wu2022conqrr,qian2022explicit,mao2023large,mao2023search,mo2023convgqr,ye2023enhancing,mo2024chiq}, which any ad-hoc search models can then use for retrieval. 
However, when involved in personalized requirements, the system build-up becomes more complex.
In real-world scenarios, the backgrounds of users are different, thus, the search results from the personalized CIR systems should yield different results for the same query according to the user profiles. For example, taking into account the environmental conditions, two farmers from different places would obtain different search results when querying the same planting guidance. 
To address this specific information need, the challenge lies in incorporating personalized context effectively to address the user search intent, e.g., producing personalized reformulated queries.
However, each user's background may contain different aspects, which might not all be related to the query in the current conversation turn. Thus, it is critical to investigate how the user information can be leveraged in query reformulation, in addition to the conversation context.

To facilitate the studies on personalized CIR, the Interactive Knowledge Assistance Track (iKAT) of Text Retrieval Conference (TREC)~\cite{aliannejadi2024trec} proposed the first experiment and provided a dataset with Personal Textual Knowledge Base (PTKB) for each conversational session. PTKB contains a set of natural language sentences to describe various characteristics or preferences of a user. However, the sentences in PTKB are not all relevant or useful for a specific query in CIR. 
To identify the relevant sentences from the PTKB for a specific query turn, an intuitive idea would be to train a classification model based on the provided annotations. Unfortunately, this approach is not feasible due to data scarcity.
Besides, relevance judgments in personalized CIR are more complex, requiring consideration of both the user's personal information and the conversation context. This complexity might lead to discrepancies in human annotations. 

In this paper, we investigate approaches to leverage personal background and conversation context for CIR without large training data. We address the following research questions: 

\textbf{RQ1}: What is the best practice for PTKB selection that can align with the relevance judgment of the retrieval task?

\textbf{RQ2}: What is the best practice to incorporate PTKB into query reformulation to improve personalized conversational IR?

\textbf{RQ3}: To what extent can the existing large language models (LLMs) consider personalized aspects while reformulating query?

To address these inquiries, we first investigate different approaches to obtain the PTKB relevance judgment, including annotated by humans, LLM, and the automatic label according to the impact of retrieval results. Then, we evaluate their effectiveness by applying the selected PTKB with the corresponding annotation for query reformulation on both sparse and dense retrieval.
Furthermore, we compare the utilization of PKTB within LLM-based query reformulation in two different settings and further design two LLM-aided strategies with in-context learning.
The experimental results show that PTKB might not always improve the retrieval performance, which might be attributed to the possible discrepancy of the annotation procedure and the lack of a proper paradigm for using PTKB. Though with the data scarcity issues, the LLM still demonstrates its potential to solve the personalized CIR problem 
when some high-quality guidance is provided.

\section{Task Definition}
Personalized conversational information retrieval aims to retrieve relevant documents $d^+$ from a large collection $D$ to satisfy the information need of the current query turn $q_n^u$ for a specific user, based on the  historical conversational context $\mathcal{H}=\{q_i\}_{i=1}^{n-1}$ and the personal text knowledge base (PTKB) $\mathcal{U}={\{s_t\}}_{t=1}^T$~\cite{aliannejadi2024trec}, where
$q_i$ and $s_t$ denote the $i$-th query turn and a ($t$-th) sentence in PTKB. 
The current query turn $q_n^u$ is context-dependent and conceals personalized information needs. Our goal is to obtain a reformulated query $q_n^{\prime}$ condition on not only the given historical conversational context $\mathcal{H}=\{q_i\}_{i=1}^{n-1}$, but also 
the associated PTKB $\mathcal{U}$.

\section{Methodology}
\subsection{Annotation Approaches Analysis}
Two types of relevance judgments evaluation are provided in the iKAT~\cite{aliannejadi2024trec}. One is between each sentence in PTKB and each query turn, indicating whether a sentence description is related to a query based on human intuitive judgment. Another one is the query-document relevance judgments used for retrieval evaluation. 

It is tempting to believe that human-annotated sentences are useful for query reformulation. However, the discrepancy might arise when sentence relevance is judged independently from retrieval effectiveness: the annotated relevant sentences may not necessarily provide useful information for improving retrieval results, as we will show in our experiments. This raises an important question: \textit{what sentence in PTKB should be leveraged for query reformulation?} We investigate two alternative approaches to determine relevant sentences in PTKB:

\textbf{(1)} Following the assumption~\cite{cao2008selecting,he2009cikm,mao2022curriculum,mo2023learning} that the relevant information in PTKB should make a positive impact when it is used for current turn query reformulation, we make automatic annotation of each sentence in PTKB based on its impact on retrieval results: if it increases retrieval effectiveness, then it is deemed relevant.

\textbf{(2)} Besides, drawing inspiration from recent successes in harnessing large language models (LLMs) as evaluators in various downstream tasks~\cite{alpaca_eval,wang2024user,huang2024c}, we also leverage LLM to obtain automatic annotations by prompting LLM to select the relevant sentences from PTKB for each query turn.
 
Therefore, we have three types of annotation to analyze the impacts of PTKB selection for the retrieval task. 
These annotation approaches are denoted as ``Human'', ``Automatic'', and ``LLM''. 

\subsection{LLM-aided Personalized Query Reformulation}
The construction of existing (personalized) CIR datasets~\cite{dalton2020trec,dalton2021cast,dalton2022cast,aliannejadi2024trec} heavily relies on human efforts, resulting in insufficient data samples to adequately support fine-tuning of end-to-end models.
We employ LLM for personalized query reformulation to alleviate such issues by considering corresponding PTKB, historical conversational context, and current turn query without specific training.

\noindent\textbf{Query Reformulation with PTKB.}
We leverage LLM to perform query reformulation with different settings, which includes providing no or all PTKB, and the selected sentences corresponding to three types of annotation approaches.
The prompt instruction template is structured as \textit{[Instruction, Input]}. The \textit{Instruction} is used to ask the LLM to generate both the rewrite and response simultaneously since the LLM-stored knowledge might help to achieve better consistency and accuracy between the generated rewrites and responses.
The \textit{Input} is composed of the selected sentences from PTKB $\mathcal{U}$, the user query $q_n^{u}$ and the conversation context $\mathcal{H}$ of the current turn $n$.

\noindent\textbf{LLM-aided Strategies.}
Then we further propose two strategies: \textit{select then reformulate} (STR) and \textit{select and reformulate} (SAR) to explicitly and implicitly incorporate PTKB for query reformulation. 
Instead of simply asking the LLM first select the relevant sentences from PTKB and then use them for query reformulation~\cite{mao2023large}, the STR first generates a hypothetical response for the current query turn by prompting the LLM with the provided whole PTKB and the conversational session, then generates the reformulated query based on the hypothetical response.
Different from STR, the SAR achieves query reformulation at once where we conjecture that implicitly performing PTKB selection at the same time could compensate for the inconsistency between the two stages.
The prompt used still follows the aforementioned template as \textit{[Instruction, Input]}, while the \textit{Instruction} is adapted with the used strategies.

\noindent\textbf{In-Context Learning.}
It is not trivial to select relevant information from PTKB for reformulating a query that can improve the personalized CIR performance since the LLM might not be able to achieve it without enough guidance.
Thus, we further leverage the in-context learning techniques to enhance the two proposed LLM-aided strategies.
Concretely, we construct a few samples from the provided training set with the annotated PTKB selection based on its impact on retrieval results, where we expect these annotated samples can guide the LLM to understand what the relevant information in PTKB for each query turn should be.
The prompt instruction template is structured as \textit{[Instruction, Demonstration, Input]}, where the \textit{Demonstration} denotes the in-context example randomly selected from the training set. The \textit{Instruction} asks the model to reformulate the current query turn conditioned on the in-context learning samples. 

\section{Experiments}
\label{sec:Experiments}
\subsection{Experimental Setup}
\noindent \textbf{Dataset and Evaluation.}
We conduct experiments on the only available dataset TREC-iKAT 2023~\cite{aliannejadi2024trec}. Different from the previous conversational IR datasets~\cite{dalton2020trec,dalton2021cast,dalton2022cast,owoicho2022trec}, in addition to the conversational sessions, it provides the Personal Textual Knowledge Base (PTKB), containing a set of natural language sentences describing the characteristics or preferences of the users. The document collection is based on ClueWeb 22B Corpus~\cite{overwijk2022clueweb22}. The statistic of the dataset is provided in Table~\ref{table: dataset}.
We use the \textit{pytrec\_eval} tool~\cite{sigir18_pytrec_eval} to compute various standard metrics: MRR, NDCG, and MAP.

\noindent \textbf{Implementation Details.} 
The dense retriever (ANCE)~\cite{xiong2020approximate} and sparse retriever (BM25) are run using Faiss~\cite{johnson2019billion} and Pyserini~\cite{lin2021pyserini}, respectively.
The different query reformulation strategies are implemented based on OpenAI's ChatGPT (gpt-turbo-3.5-16k) API with the default hyper-parameters. The complete prompt template and more details can be found in our released code.\footnote{\url{https://github.com/fengranMark/PersonalizedCIR}}

\begin{table}[!t]
\centering
\caption{Statistics of TREC-iKAT 2023 dataset.}
\vspace{-3ex}
\begin{tabular}{lll}
\toprule
\textbf{TREC-iKAT} & \textbf{Train} & \textbf{Test} \\ 
\midrule
\# Topic &8 & 13 \\
\# Conversations &11 & 25 \\
\# Turns (Queries) & 95 & 324 \\\# Assessed Turns (Queries) & - & 176 \\
\# PTKB assessed turns &42 & 112 \\
\# PTKB assessments &368 & 1,158 \\
\# Relevant PTKB &64 & 182 \\
\# Collection & \multicolumn{2}{c}{11.6M} \\
\bottomrule
\end{tabular}
\label{table: dataset}
\vspace{-2ex}
\end{table}

\subsection{Analysis of PTKB Annotation}
We first address the \textbf{RQ1} by analyzing the statistics and effectiveness of different annotation approaches. 
The identification of whether a query turn needs a PTKB sentence is based on whether it leads to any improvement in search results when incorporated in query reformulation.

\subsubsection{Annotation Statistic}
\begin{figure}[!t]
\centering
\includegraphics[width=0.45\textwidth]{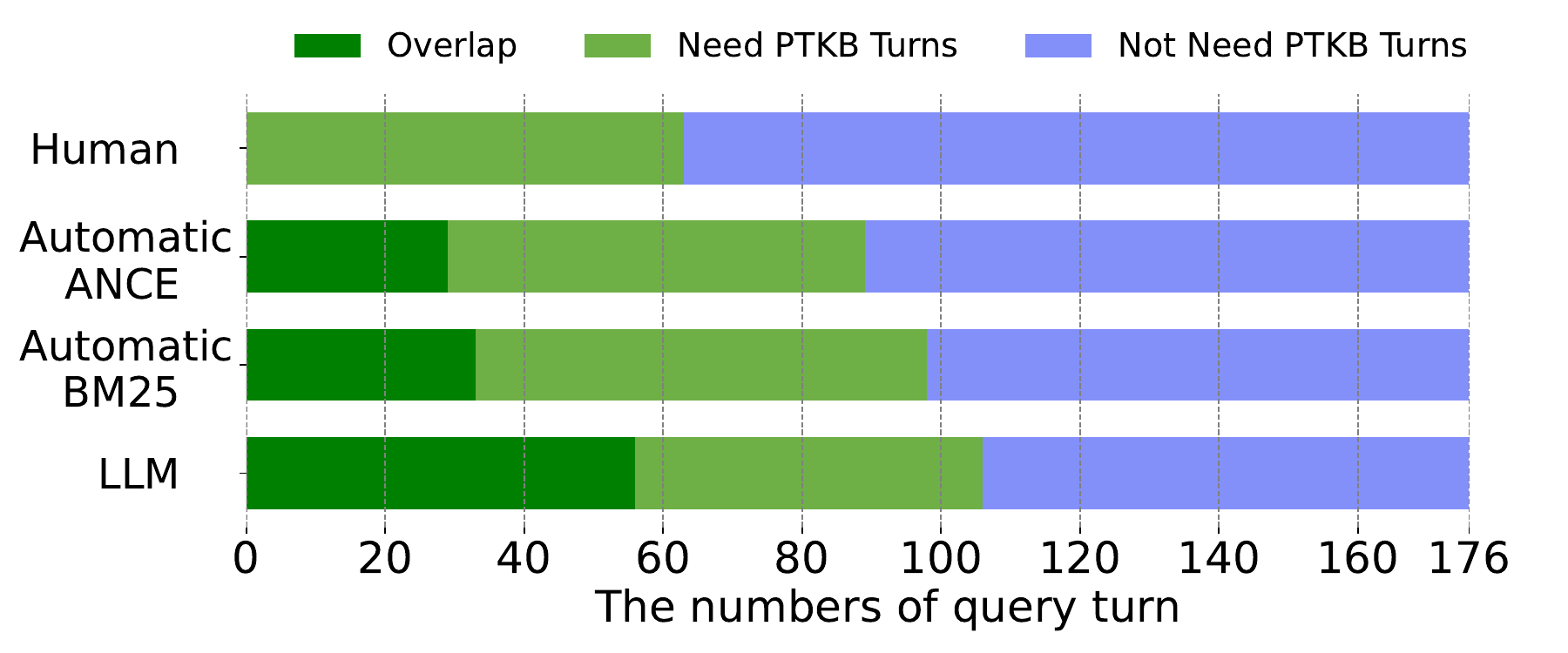}
    \caption{The statistic of three different types of PTKB annotation. The ``overlap'' denotes the same annotation as Human.}
\label{fig: overlap}
\vspace{-3ex}
\end{figure}

Figure~\ref{fig: overlap} shows the results of labeling the relevance between each query turn with PTKB. 
We observe a big difference across the annotation approaches, e.g., the overlap with human annotation is small, which implies the annotation based on human intuition might not always align with the impact of search results (Automatic) and the judgment of LLM.
Besides, the judgment of the necessity of PTKB shows a certain degree of disagreement in each annotation approach, which indicates the difficulty of identifying relevant information from PTKB for a query. 

\subsubsection{Effectiveness}
\begin{figure}[!t]
\centering
\includegraphics[width=0.4\textwidth]{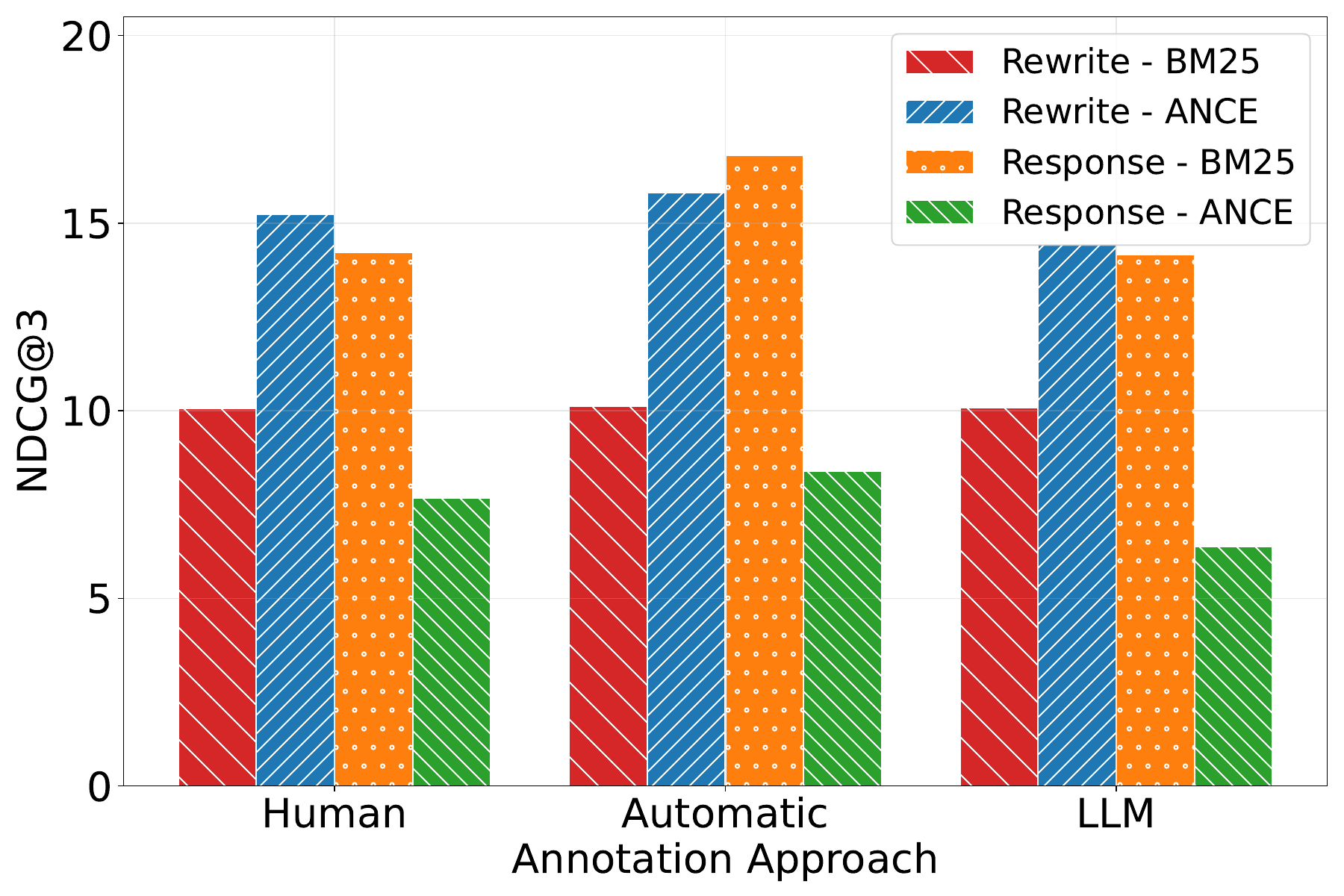}
    \caption{The results of three types of annotation approach with different reformulated query forms and retrievers.}
\label{fig: Annotation}
\vspace{-3ex}
\end{figure}

The retrieval performance of different annotation approaches is shown in Figure~\ref{fig: Annotation}, which includes using two query forms and retrievers. We can see the automatic annotation produced based on the impact of retrieval results obtains better performance than the others, which demonstrates the advantage of aligning sentence relevance annotations in PTKB and retrieval effectiveness.
In addition, we can see that BM25 prefers using the response as the search query while the ANCE prefers to rewrite. 
This might be due to much more lexical matches by BM25 in the generated response, while the topic drift phenomenon happens in the semantic match in ANCE. 
Following this observation, we use the concatenation of rewrite and response for the BM25 and the rewrite only for the ANCE in the remaining experiments.

\begin{table}[t]
\centering
\caption{The effectiveness of different PTKB incorporation methods. Bold and $\dagger$ denote the best results and significant improvements with t-test at $p<0.05$ over the remaining compared results on each evaluation setting.}
\vspace{-2ex}
\begin{tabular}{lccccc}
\toprule
\textbf{Model} & \textbf{Method} & \textbf{MRR} & \textbf{N@3} & \textbf{N@5} & \textbf{MAP}\\
\midrule
\multicolumn{6}{c}{\cellcolor[HTML]{fff8f8}{Evaluate on the whole test set (176 turns)}} \\
\midrule
\multirow{6}{*}{BM25} & None & \textbf{44.35}$^\dagger$ & \textbf{21.22}$^\dagger$ & \textbf{20.68}$^\dagger$ & \textbf{8.91} \\
~ & Use all & 40.36 & 19.19 & 18.84 & 8.28  \\
~ & Human & 41.65 &19.66 & 19.46 &8.82\\
~ & Automatic & 40.29 & 19.12 & 18.87 & 8.58 \\
~ & STR & 41.53 & 18.96 & 18.09 & 8.37 \\
~ & SAR & 36.04 & 17.48 & 16.87 & 8.02 \\
\midrule
\multirow{6}{*}{ANCE} & None & 32.47 & 14.25 & 13.73 & 5.68\\
~ & Use all & 33.64 & 15.30 & 15.09 & 6.13 \\
~ & Human & 33.63 & 15.98 & 15.69 & 6.16\\
~ & Automatic & 31.08 & 14.36 & 14.01 & 5.89 \\
~ & STR &32.37 &15.05 &14.02 &5.72 \\
~ & SAR &31.76 &14.78 &15.12 &5.47 \\
\midrule
\multicolumn{6}{c}{\cellcolor[HTML]{fff8f8}{Evaluate on the subset of need PTKB (67 turns)}} \\
\midrule
\multirow{6}{*}{BM25} & None & 31.47 & 15.40 & 14.75 & 4.71 \\
~ & Use all & 29.17 & 13.16 & 12.66 & 4.05 \\
~ & Human & 33.08 & 14.48 & 14.40 & 4.74 \\
~ & Automatic & \textbf{33.72}$^\dagger$ & \textbf{16.78}$^\dagger$ & \textbf{16.28}$^\dagger$ & \textbf{5.33}$^\dagger$ \\
~ & STR & 32.38 & 13.66 & 13.08 & 4.83 \\
~ & SAR & 31.56 & 13.34 & 11.36 & 3.90 \\
\midrule
\multirow{6}{*}{ANCE} & None &21.35 & 7.90 &7.45 & 2.63 \\
~ & Use all &23.45 & 10.82 &9.88 & 3.18 \\
~ & Human &25.54 & 10.57 &10.33 & 3.58 \\
~ & Automatic &25.19 & 10.74 &10.23 & 3.74 \\
~ & STR &25.45 & 10.25 &10.12 & 3.31 \\
~ & SAR & 24.88 & 10.85 & 9.77 & 3.68 \\
\bottomrule
\end{tabular}
\label{table: main results}
\vspace{-4ex}
\end{table}

\begin{table}[t]
\centering
\caption{The effectiveness of 
enhancing two LLM-aided strategies with few-shot in-context learning. Bold indicates the best results on each retrieval type. $\dagger$ denotes significant improvements with t-test at $p<0.05$ over the best performance on the whole test set evaluation in Table~\ref{table: main results}.}
\vspace{-2ex}
\resizebox{\columnwidth}{!}{
\begin{tabular}{lcccccc}
\toprule
\textbf{Model} & \textbf{Method} & \textbf{Sample} & \textbf{MRR} & \textbf{N@3} & \textbf{N@5} & \textbf{MAP}\\
\midrule
\multirow{8}{*}{BM25} & \multirow{4}{*}{STR} & 0-shot & 41.53 & 18.96 & 18.09 & 8.37 \\
~ & ~ & 1-shot & 41.10 & 19.61 & 19.94 & 8.56 \\
~ & ~ & 3-shot & 43.41 & 20.76 & 20.43 & 8.40 \\
~ & ~ & 5-shot & 41.54 & 19.08 & 18.81 & 8.53 \\
\cmidrule(lr){2-7}
~ & \multirow{4}{*}{SAR} & 0-shot & 36.04 & 17.48 & 16.87 & 8.02 \\
~ & ~ & 1-shot & 39.55 & 17.80 & 18.01 & 8.80\\
~ & ~ & 3-shot & \textbf{45.73}$^\dagger$ & \textbf{22.72}$^\dagger$ & \textbf{22.01}$^\dagger$ & \textbf{10.37}$^\dagger$\\
~ & ~ & 5-shot & 44.28 & 22.01 & 21.61$^\dagger$ & 10.27$^\dagger$\\
\midrule
\multirow{8}{*}{ANCE} & \multirow{4}{*}{STR} & 0-shot & 31.76 & 17.78 & 15.12 & 5.47 \\
~ & ~ & 1-shot & 27.66 & 12.89 & 12.16 & 5.44 \\
~ & ~ & 3-shot & 33.49 & 15.34 & 13.93 & 5.94 \\
~ & ~ & 5-shot & 28.61 & 13.18 & 12.63 & 5.16\\
\cmidrule(lr){2-7}
~ & \multirow{4}{*}{SAR} & 0-shot & 32.37 & 15.05 & 14.02 & 5.72 \\
~ & ~ & 1-shot & 32.49 & 14.45 & 13.52 & 5.69 \\
~ & ~ & 3-shot & \textbf{38.27} & \textbf{17.35} & \textbf{17.37} & \textbf{6.83} \\
~ & ~ & 5-shot & 33.13 & 16.03 & 15.38 & 6.26 \\
\bottomrule
\end{tabular}}
\label{table: in-context}
\vspace{-4ex}
\end{table}

\subsection{Main Results}
In this section, we address the \textbf{RQ2} and the \textbf{RQ3} by incorporating PTKB for query reformulation with different strategies and evaluate both the whole test set and a subset of only contain the turns that need PTKB for improvement. 
The full set evaluation aims to provide a comprehensive understanding of the performance of existing general techniques with various information usage, while the subset evaluation tries to show results in an exact personalized CIR scenario.
The zero-shot results and the in-context learning performance are reported in Table~\ref{table: main results} and Table~\ref{table: in-context}. 

\subsubsection{Comparison of Different Strategies}
From Table~\ref{table: main results}, we can find BM25 performs better than ANCE consistently, which is contrary to previous TREC-CAsT (CIR) datasets evaluation~\cite{lin2021contextualized,mao2023learning}. This might be due to the decision-originated query in this dataset~\cite{aliannejadi2024trec} that contains much more lexical matches.

When evaluated on the whole test set, it is striking to observe that the best strategy is not using any PTKB information. 
This implies that the query turns do not need to be all personalized. This is consistent with earlier studies on search personalization~\cite{dou2007large,teevan2008personalize}, which show variable effects of personalization on different queries.
However, within the subset of queries that need PTKB, the automatic annotation achieves the best results, confirming our assumption that the relevance of PTKB sentences should be associated with retrieval effectiveness.
We find that leveraging PTKB can improve the retrieval performance compared with not using it in this evaluation subset, since personalized search might not be necessarily uniformly applied to all users and queries in practice, a mechanism for identifying whether a query turn requests personalized information needs is desirable and we leave this as future work. 

In addition, we also observe that not using any PTKB performs better than simply using all within the BM25 evaluation, which suggests the LLM could somehow implicitly consider the personalized requirements. Using the two PTKB selection strategies STR and SAR can also improve the performance against using no and all PTKB in both sparse and dense retrieval, suggesting a better strategy for the usage of LLM is desirable.

\subsubsection{In-Context Learning Performance}
From Table~\ref{table: in-context}, we can find BM25 still performs better than ANCE as our previous observation. Compared with zero-shot runs, the few-shot samples can provide the context to guide LLM to learn how to identify relevant PTKB to achieve better query reformulation results in most cases. It also outperforms the best results in Table~\ref{table: main results}, which confirms the effectiveness of deploying in-context learning with LLM under the data scarcity scenario.

Among the few-shot methods, three-shot achieves the best results, which indicates we should focus on the quality rather than the quantity of the used samples.
Besides, different from the observation in Table~\ref{table: main results}, we can find that SAR performs better than STR with more in-context samples. Such phenomenons indicate simultaneously conducting relevant PTKB selection and personalized query reformulation might implicitly help each other when enough in-context guidance is provided to LLM.

\section{Conclusion and Future Work}
In this paper, we explore the feasibility of incorporating the personal textual knowledge base (PTKB) for personalized conversational information retrieval based on LLM. We find the potential discrepancy in the existing human relevance judgment annotation of PTKB sentences with respect to retrieval effectiveness. We thus develop two alternative approaches to leverage PTKB, by automatically annotating sentences according to their impact on retrieval and by prompting an LLM for query reformulation. Our experiments confirm that the automatic annotation aligning sentence relevant with retrieval effectiveness is a better approach than human annotations. We also demonstrate that the LLM is a powerful tool to connect PTKB selection and query reformulation, especially when a few high-quality examples are provided.

The problem of leveraging PTKB is crucial for personalized CIR. We focused on selecting relevant sentences in this paper for all queries as preliminary research. Selective personalization for CIR should be investigated in the future. PTKB can also be exploited from a more sophisticated user modeling perspective rather than merely being seen as providing a set of sentences.


\bibliographystyle{ACM-Reference-Format}
\bibliography{sample-base}

\appendix

\end{document}